\documentstyle[preprint,eqsecnum,aps]{revtex}

\begin{document}
\draft
\tightenlines
\preprint{MSUCL-926}
\title{Photons from axial-vector radiative decay in a hadron gas}
\author{Kevin Haglin\cite{myemail}}
\address{
National Superconducting Cyclotron Laboratory, Michigan
State University\\
East Lansing, Michigan 48824--1321
}
\date{\today}
\maketitle
\begin{abstract}
Strange and non-strange axial-vector meson radiative decays contribute
to photon production in hadron gas. One- and two-hadron radiative
decay modes of  $b_{1}(1235)$, $a_{1}(1260)$ and $K_{1}(1270)$ are
studied.  At 200 MeV temperature and for a narrow range in photon
energies they contribute more to the net thermal
photon production rate than $\pi\rho\rightarrow \pi\gamma$,
$\pi\pi\rightarrow \rho\gamma$ or $\rho\rightarrow\pi\pi\gamma$.
They provide significant contribution to the rate for photon
energies as high as 1.5--2.0 GeV.  For higher energies they
are less important.
\end{abstract}

\narrowtext

\section{Introduction}
\label{sec:intro}

Photons are able to probe any or all stages of
a relativistic heavy-ion collision\cite{es80} since their
mean free paths are much larger than the transverse size of
the hot and dense region and will therefore escape after production
without rescattering.  Massive photons, or dileptons, share
this property.  Therefore, photons and dileptons
are thought to provide exciting means of probing even the
central region cleanly\cite{elf76,es78,kk81,fh82,bs83,rh85,gs86}.
Observation of quark-gluon plasma through
the use of photons is contingent, among other things, on the
production rate or ultimately the yield being distinguishable
from hadron gas.  Yet, current understanding is that
hadron gas and quark-gluon plasma produce energetic
photons nearly equally frequently at fixed temperature
$\approx 200$ MeV\cite{jkplds}.  While studying low-mass
dilepton spectra after integrating the rates over
space-time history the same conclusion was reached: the two
phases produced massive photons more or less equally\cite{khcgve}.
If any important contribution has been ignored, e.g. strange
particles or heavy mesons, it would be very useful to know about
such things.

Within the hadron gas photon production will originate in many
ways as there may be several hadronic species produced directly
or perhaps pair-produced thermally since the temperatures can
rise to extreme values and approach the critical value $T_{c}$, somewhere
in the range 150--200 MeV, where crossover to the deconfined and
chirally symmetric phase is presumed to be.  Produced photons might
be categorized according to their energy as follows.
For $E_{\gamma}$ below 0.5 GeV bremsstrahlung from
charged particle scattering and decay will be important.
Also, reactions in which a large mass hadron (compared to the
combined mass of the initial hadrons) is in the final state
such as $\pi\pi\rightarrow\rho\gamma$ are endothermic and tend to
give large contributions at low photon energies.
For higher energies direct decays give a substantial contribution
as does the process $\pi\rho\rightarrow \pi\gamma$.
It has recently been shown that inclusion of the $a_{1}$ resonance
in $\pi\rho\rightarrow \pi\gamma$ scattering results in a
large rate of photon production, larger than all others combined\cite{shuryak}.
When interference effects are treated properly the effect of
the $a_{1}$ is changed somewhat but remains important\cite{song93}.
Vector meson decays are also important: the $\omega$ has a
relatively large decay rate into a neutral pion and photon.  Folding
in a Bose-Einstein distribution for the $\omega$ results in
a significant thermal production rate~\cite{jkplds}.
Kaon abundances are 2--3 times smaller than pion's
at temperatures near $T_{c}$, so they will frequently scatter with pions and
with rhos.  With the pions, $K^*$(892) resonances can be formed and
with the rhos, scattering through $K_{1}(1270)$ is probable.
The $K\rho$ decay channel of $K_{1}(1270)$ is exceptional in that
it is only 71 MeV/c
center-of-mass momentum above threshold and therefore has
a cross section (proportional to $1/{\bbox{p}}_{\rm c.m.}^{2}$) which is
relatively large.  This resonance has recently been shown to affect
kaon mean free paths in hot matter by tens of percents\cite{khsp}.  It
is therefore interesting to study reactions involving the $K_{1}$ and
their contributions to photon production.

The large family of non-strange and strange mesons will
comprise the hot thermalized system.  At temperatures near 100 MeV or so,
the system will be populated mostly by pions since
Boltzmann weightings strongly suppress the heavier ones.  As the temperature
rises to something near $T_{c}$, the situation can become different
since the suppression is effectively weaker.
Using zero chemical potentials in Boltzmann distributions at a temperature
of 200 MeV, there are $2.9\times 10^{-1}$ $\pi$\, s,
$1.5\times 10^{-1}$ $\rho$\,s and $1.3\times 10^{-1}$ $K^*$\,s
per fm$^{3}$, while there are $2.9\times 10^{-2}$ $K_{1}(1270)$\,s and
$2.6\times 10^{-2}$ $b_{1}(1235)$\,s per fm$^{3}$\cite{khsp}.
The latter is quite important because it decays predominantly
into a $\pi\omega$ combination and has a full width of 155 $\pm 8$ MeV.
The $\omega$ might be on shell in which case it just contributes to the
equilibrium number of omegas.  It might also be off shell and
will subsequently decay (off shell) into $\pi^{0}\gamma$.  The product
of the density times overall decay seems sizeable, so folding in the
dynamics, kinematics and thermally distributed phase space could
result in a significant thermal production rate.

Theoretical framework for photon production rates through
radiative decay from a thermalized system will be presented
in the next section.  Discussion of the model
dependence of hadronic and electromagnetic transitions upon their
specific interactions will also appear.
They are modelled by effective Lagrangians describing
both $AV\phi$ and $VV^{\prime}\phi$
three-point vertices, where $A$ is an axial-vector field, $V$ and
$V^{\prime}$ are vector fields and $\phi$ is a pseudoscalar field.
Then in section \ref{sec:results} results are presented for
one- and two-hadron radiative decays for a temperature 200 MeV.
The channel $b_{1}\rightarrow \pi\pi^{0}\gamma$ is found to be important,
and so further dependence on temperature is studied.  Section
\ref{sec:broadening} contains an estimate for the collisional broadening
of $\omega$ in the above mentioned $b_{1}$ radiative
decay.  Modified photon production rates follow.
Finally, section \ref{sec:conclusion} contains
concluding remarks and possible ramifications of these results.

\section{Thermal radiative decay}
\label{sec:general_theory}

The rate for photon production from a thermalized system at temperature
$T$ whose size is small relative to the photon mean free path is
proportional to the imaginary part of the retarded photon
self-energy $\Pi^{\mu\nu}_{R}$ and a thermal weighting
as\cite{weldon83,lmtt85,cgjk91}
\begin{equation}
E_{\gamma}{dR\over d^{3}p_{\gamma}} = {-2g^{\mu\nu}\over (2\pi)^{3}}
{\rm Im}\Pi^{R}_{\mu\nu} {1\over e^{E_{\gamma}/T}-1}.
\end{equation}
For temperatures $100< T < T_{c}$, the largest contributions to
the self-energy will
be one- and two-loop diagrams consisting of $\pi$\,s and $\rho$\,s.
Near $T_{c}$, contributions
from diagrams in which heavier non-strange and strange particles
occupy the loops also become important.
Figure~\ref{fig:self-energy}a shows one-loop $\pi a_{1}$ and $K K_{1}$
contributions, and \ref{fig:self-energy}b shows a two-loop $\pi \omega$
contribution [with $\omega$ further splitting to $b_{1}\pi$].  If
the imaginary part of any of these diagrams (obtained by cutting them)
gives a calculated width in vacuum that is relatively sizeable, then
even being less abundant than pions or rhos in the hot system, it is
interesting to ask about their contributions
to photon production.  Cutting the diagram in Fig.~\ref{fig:self-energy}a
results in single-hadron radiative decays of either $a_{1}$ or
$K_{1}$ axial-vectors.  They are shown in Fig.~\ref{fig:decays}a.
The diagram in Fig.~\ref{fig:self-energy}b is of two-loop order and therefore
its cut diagram has four external lines.  Kinematics allows two
possibilities for the photon being in the final state.  First, there could
be $\pi b_{1}\rightarrow \omega \rightarrow \pi^{0}\gamma$ scattering.
Secondly, a two-pion radiative decay
$b_{1}\rightarrow \omega\pi\rightarrow \pi\pi^{0}\gamma$ can proceed.
The production rate from the first process turns out
to be rather unimportant as compared with the second.
Its diagram is in Fig.~\ref{fig:decays}b.

Elementary radiative hadron decays,
$h_{a} \rightarrow \sum_{b} h_{b} + \gamma$, can proceed
in two ways.  One of the incoming or outgoing particles
can emit a photon (bremsstrahlung) or secondly, the photon can
be emitted directly.  Direct emission reflects the internal structure
and has vanishing phase space for vanishing photon energy.
In a thermal
calculation the decay rate is important but so too
is the number density of the decaying hadron.  Besides
$\rho\rightarrow \pi\pi\gamma$, the many-hadron final states have
been neglected since they were assumed
small.  But as will soon be seen, at least one is not.
The thermal rate for radiative hadronic decay
is
\begin{eqnarray}
E_{\gamma}{dR\over d^{3}p_{\gamma}} &=& {\cal N}\int
{d^{3}p_{a}\over (2\pi)^{3}2E_{a}}
f(E_{a})
|{\cal M}|^{2} (2\pi)^{4} \delta^{4}(p_{a}-p_{1}-\ldots -p_{n}-p_{\gamma})
\nonumber\\
& & \times\left\lbrace\prod\limits_{i=1}^{n}{d^{3}p_{i}\over (2\pi)^{3}2E_{i}}
[1\pm f(E_{i})]\right\rbrace {1\over (2\pi)^{3}2}.
\label{eq:decayrate}
\end{eqnarray}
where ${\cal N}$ is the degeneracy and $f(E)$ is either a Bose-Einstein
or Fermi-Dirac distribution depending on the species.  Bose
enhancement(Pauli suppression) is enforced by choosing the $+(-)$ sign
in the square-bracketed term.
Identifying the species of hadrons and their interaction with
other hadrons into which they might decay, and specifying their
interaction with the electromagnetic field completely defines the
problem.  What remains is to carry out the necessary four-vector
algebra and phase space integration.

The model Lagrangians used are the
following.  First for the $AV\phi$ interaction\cite{shuryak}
\begin{equation}
L_{AV\phi} = g\, \left. A^{\mu}\right((p_{V}\cdot p_{\phi})g_{\mu\nu}
-p_{V\, \mu}p_{\phi \, \nu}\left) V^{\nu} \right. \phi ,
\label{eq:lavphi}
\end{equation}
and then for the $VV^{\prime}\phi$ vertex\cite{um88}
\begin{eqnarray}
L_{VV^{\prime}\phi} &=& g^{\prime}\,
\epsilon_{\mu\nu\alpha\beta}\, p_{V}^{\mu}\, V^{\nu}
\, p_{V^{\prime}}^{\alpha}\, V^{\prime\, \beta}\, \phi
\label{eq:lvvphi}
\end{eqnarray}
where $\epsilon_{\mu\nu\alpha\beta}$ is the totally antisymmetric
unit tensor.  Modelling the interactions depicted by vertices in
the diagrams of Figs.~\ref{fig:decays}a and b this way, their
individual decay widths are
calculated to be
\begin{equation}
\Gamma_{\omega\rightarrow \pi^{0}\gamma} =
{g_{\omega\pi^{0}\gamma}^{2} \over 12\pi m_{\omega}^{2}} |{\bbox{p}}\/|
\left(p_{\pi^{0}}\cdot p_{\gamma}\right)^{2}
\end{equation}
\begin{eqnarray}
\Gamma_{b_{1}\rightarrow \pi\omega} &=&
{g_{b_{1}\pi\omega}^{2} \over 24\pi m_{b_{1}}^{2}} |{\bbox{p}}\/|
\left(2(p_{\pi}\cdot p_{\omega})^{2}+m_{\omega}^{2}(m_{\pi}^{2}
+{\bbox{p}}^{2})\right) \\
\Gamma_{a_{1}\rightarrow \pi\rho} &=&
{g_{a_{1}\pi\rho}^{2} \over 24\pi m_{a_{1}}^{2}} |{\bbox{p}}\/|
\left(2(p_{\pi}\cdot p_{\rho})^{2}+m_{\rho}^{2}(m_{\pi}^{2}
+{\bbox{p}}^{2})\right)
\end{eqnarray}
and
\begin{equation}
\Gamma_{K_{1}\rightarrow \rho K} = {g_{K_{1}\rho K}^{2} \over 24\pi
m_{K_{1}}^{2}} |{\bbox{p}}\/| \left( 2(p_{K}\cdot p_{\rho})^{2}+
m_{\rho}^{2}(m_{K}^{2}+{\bbox{p}}^{2})\right),
\end{equation}
where ${\bbox{p}}$ is the center-of-mass momentum of the decay products.
These determine the coupling constants
to be $g_{\omega \pi^{0}\gamma}=0.7$,
$g_{b_{1} \pi\omega}=10.3$,
$g_{a_{1} \pi\rho}=14.8$ and
$g_{K_{1}\rho K} = 12.0$ GeV$^{-1}$ in order that the
partial decay widths are 0.7, 155, 400 and 37.8 MeV respectively,
so they match results from the Review of Particle
Properties\cite{pdg}.

An estimate for coupling the axial-vectors $a_{1}$ and $K_{1}$
to the photon field (and a pion or kaon respectively) is now needed.
Decay of the $b_{1}$ into a pion
and photon is not considered since it results in a small rate.
Here is where a model is employed since no
experimental information on these decays
exist.  By vector-meson dominance, the coupling for
these are
\begin{eqnarray}
g_{a_{1}\pi\gamma} &=& (e/f_{\rho})g_{a_{1}\rho\pi}\nonumber\\
g_{K_{1}K\gamma} &=& (e/f_{\rho})g_{K_{1}\rho K}
\end{eqnarray}
where $f_{\rho}^{2}/4\pi$ = 2.9 is the
$\rho \pi\pi$ coupling.  Numerically
these coupling constants become $g_{a_{1}\gamma\pi}$ = 0.74 GeV$^{-1}$ and
$g_{K_{1}\gamma K}$ = 0.60 GeV$^{-1}$.
Using the calculated expressions
\begin{eqnarray}
\Gamma_{a_{1}\rightarrow \gamma \pi} &=& {g_{a_{1}\gamma\pi}^{2} \over 12\pi
m_{a_{1}}^{2}} |{\bbox{p}}\/| \left(p_{\gamma}\cdot p_{\pi}\right)^{2}\\
\Gamma_{K_{1}\rightarrow \gamma K} &=& {g_{K_{1}\gamma K}^{2} \over 12\pi
m_{K_{1}}^{2}} |{\bbox{p}}\/| \left(p_{\gamma}\cdot p_{K}\right)^{2},
\end{eqnarray}
electromagnetic decay widths of $\Gamma_{a_{1}\rightarrow \gamma \pi}$ =1.4 MeV
and $\Gamma_{K_{1}\rightarrow \gamma K}$ = 1.5 MeV are obtained.

Matrix elements for photon production corresponding to the processes
in Fig.~\ref{fig:decays}a for $a_{1}$ and $K_{1}$ decay are
\begin{eqnarray}
{\cal M} &=& g_{a_{1}\gamma \pi}\, \left. \epsilon_{a_{1}}^{\mu}\,
\right((p_{\pi}\cdot p_{\gamma})g_{\mu\nu} - p_{\gamma \, \mu}p_{\pi \, \nu}
\left) \, \epsilon_{\gamma}^{\nu} \right.
\end{eqnarray}
and
\begin{eqnarray}
{\cal M} &=& g_{K_{1}\gamma K}\, \left. \epsilon_{K_{1}}^{\mu}\,
\right((p_{K}\cdot p_{\gamma})g_{\mu\nu} - p_{\gamma \, \mu}p_{K \, \nu}
\left) \, \epsilon_{\gamma}^{\nu}, \right.
\end{eqnarray}
where the $\epsilon\,$s are polarization vectors for the respective
(vector) fields.  Each one depends therefore on a spin index which
is not explicitly written.  The two-pion radiative decay depicted
in Fig.~\ref{fig:decays}b has a matrix element of
\begin{eqnarray}
{\cal M} &=& g_{b_{1}\pi\omega}\, \, g_{\omega\pi^{0}\gamma}\,
\left. \epsilon_{b_{1}}^{\mu}\,\right((p_{\pi}\cdot p_{\omega})g_{\mu\nu}
-p_{\omega \, \mu}p_{\pi \, \nu}\left) D^{\nu\alpha}\,
\epsilon_{\beta\alpha\sigma\lambda}\,
p_{\omega}^{\beta}\, p_{\gamma}^{\sigma}\, \epsilon_{\gamma}^{\lambda} \right.
\label{eq:mb1decay}
\end{eqnarray}
where the pion momenta refer to the pion emitted from the
$b_{1}\omega\pi$ vertex in the diagram of Fig.~\ref{fig:decays}b and
the propagator for the $\omega$ in the same diagram
is
\begin{equation}
D^{\nu\alpha}(l) = (g^{\nu\alpha} - l^{\nu}l^{\alpha}/m_{\omega}^{2})
{1\over l^{2}-m_{\omega}^{2}-im_{\omega}\Gamma_{\omega}}.
\label{eq:omegaprop}
\end{equation}
The width is taken as the vacuum value of $\Gamma_{\omega}$= 8.43 MeV,
although modification due to the presence of matter will be discussed
later.  Squaring the matrix elements, contracting all the indices and summing
over spin states is the first step.
Using Eq.~(\ref{eq:decayrate}) the production rate reduces to an
integral over one- and two-particle phase space respectively, for
processes from Figs.~\ref{fig:decays}a and b.

\section{Results}
\label{sec:results}

Thermal photon production rates at 200 MeV temperature
for the processes $a_{1}\rightarrow \pi\gamma$,
$K_{1}\rightarrow K\gamma$ and $b_{1}\rightarrow \pi\pi^{0}\gamma$ are
shown in Fig.~\ref{fig:three}.  Results for $a_{1}\rightarrow \pi\pi\gamma$
and $K_{1}\rightarrow K\pi\gamma$ will not be discussed here since they
result in smaller rates.  Features common to these processes are
that they turn over and approach zero for $E_{\gamma}\rightarrow 0$
simply due to vanishing phase space in this limit.  Unlike processes
such as $\rho\rightarrow \pi\pi\gamma$, these axial-vector {\em direct}
decays cannot proceed without the
photon present.  They peak at slightly different photon
energies due to the differences in the axial-vector and
decay particle masses.  Their slopes also reflect these differences:
the $K\gamma$ shows the largest slope, then the $\pi\pi^{0}\gamma$
followed by the $\pi\gamma$ final state results.  The
most startling feature is the overall magnitude of the
$b_{1}$ decay.  For this temperature its peak is seven times larger
than the others.  The reason for this is twofold.  First
is the relative abundance of $b_{1}$,
for which there are roughly half as many $b_{1}$\,s as there are
omegas per unit volume.  This is larger than one would expect,
but isospin degeneracy and the high temperature are responsible.
Secondly, $b_{1}$ decays predominantly into a $\pi\omega$ combination,
which subsequently decays into $\pi\pi^{0}\gamma$.  Roughly speaking,
the overall rate for this is
\begin{equation}
{\Gamma_{b_{1}\rightarrow\pi\omega}
\Gamma_{\omega\rightarrow\pi^{0}\gamma} \over
\Gamma_{\omega}^{\rm full}} = 13.2 \ {\rm MeV},
\label{eq:roughrate}
\end{equation}
which is a factor 18 larger than simple $\omega\rightarrow \pi^{0}\gamma$
decay.  Multiplying the density of $b_{1}$ mesons times the rate from
Eq.~(\ref{eq:roughrate}) results in a factor 9 more than radiative
$\omega$ decay.  When comparing thermal photon production via
$b_{1}\rightarrow \pi\pi^0\gamma$ with the corresponding thermal production
rate for $\omega \rightarrow \pi^{0}\gamma$ presented at $T$ = 200 MeV
in Ref.\cite{jkplds}, this factor of 9 is consistent.
Kinematics and Bose-Einstein distributions complicate matters
as does the phase space integration, but the result  can be understood
with the above simple argument.

One-pion radiative decay of real omegas is not included within
the $b_{1}$ decay---they really are separate contributions.
Imagine $b_{1}$ being instead very massive.  Thermal
photon production via $b_{1}$ decay would apprach zero in the
infinite-mass limit simply due to vanishing equilibrium number density
$\bar{n}_{b_{1}}\rightarrow 0$.  Omega radiative
decay, on the other hand, is
not affected by such a change and $\omega$
decay must contribute.  Summing both processes does not amount to
double counting.

Radiative decay of $b_{1}$ turns out to be rather large
(for a limited range in photon energy) due in part
to the abundance.  The natural question to ask therefore, is about
its rise with increasing temperature; or alternatively, its fall with
decreasing temperature.  For this reason
Fig.~\ref{fig:four} is shown at three values of temperature 100, 150
and 200 MeV.  Each result is superimposed on the rate of photon production
from $\pi\rho\rightarrow \pi\gamma$ as calculated using an effective
chiral Lagrangian including effects of the $a_{1}$ resonance
coherently\cite{song93}.   As the temperature drops, the photon
energy for which the two processes are equal shifts downward.
But the noteworthy feature of dominance of the
two-pion radiative decay of $b_{1}$ for a narrow range of
photons energies basically remains even at temperature 100 MeV.

\section{Collisional broadening of omega}
\label{sec:broadening}

The hadron-gas environment is quite different from free space, and
one can therefore justifiably question the applicability of the
matrix element of Eq.~(\ref{eq:mb1decay}).  In particular, the $\omega$
propagator is just taken to be the free space vector propagator which
contains the vacuum width of 8.43 MeV and no shift in pole position.
The vacuum width corresponds to a lifetime of 23 fm/c.  Yet, the mean free
path of an $\omega$ in this hot matter is at most a few fermis since
it can scatter with pions to form a $b_{1}$ resonance\cite{khsp3}, so it
will likely rescatter before decaying.  To account for this, a
collisional broadened width is computed.  Roughly it is $n\sigma v$, where
$n$ is the pion density, $\sigma$ is the $\pi\omega$ cross section,
and $v$ is their relative velocity.  If there are 0.3 pions per fm$^{3}$,
if the cross section is 1 fm$^{2}$ and the relative velocity $v/c$ = 0.5, an
extra {\em collisional} width of 30 MeV should be added to the
vacuum width.  Rather than use this crude estimate, the
expression
\begin{equation}
\Gamma^{\rm coll}_{\omega}(E_{\omega}) =
\int\, ds \, {d^{3}p_{\pi}\over (2\pi)^{3}}
f(E_{\pi}) \sigma_{\pi\omega}(s) v_{rel} \delta\left(s-(p_{\pi}
+p_{\omega})^{2}\right)
\end{equation}
is used, where
\begin{equation}
v_{rel} = {\sqrt{(p_{\pi}\cdot p_{\omega})^{2}-4m_{\pi}^{2}m_{\omega}^{2}}
\over E_{\pi}E_{\omega}}.
\end{equation}
A Breit-Wigner form for the cross section
\begin{equation}
\sigma_{\pi\omega}(\sqrt{s}) = {\pi \over {\bbox{k}}^{2}}
{\Gamma_{b_{1}\rightarrow \pi\omega}^{2} \over (\sqrt{s}-m_{b_{1}})^{2}
+\Gamma_{b_{1}}^{2}/4}
\end{equation}
is used with ${\bbox{k}}$ being the center-of-mass momentum and the
full and partial widths taken to be 155 MeV.
The collision rate (or width) is presented in Fig.~\ref{fig:collwidth}
for 100, 150 and 200 MeV temperature.  Energy dependence aside, it is
indeed of order 30 MeV. It is
now added to the vacuum width resulting
in an energy-dependent full width of
\begin{equation}
\Gamma_{\omega}^{\rm full}(E_{\omega}) = \Gamma_{\omega}^{\rm vac} +
\Gamma_{\omega}^{\rm coll}(E_{\omega}).
\end{equation}

The propagator of Eq.~(\ref{eq:omegaprop}) is now modified by
replacing the omega width by this energy-dependent full width $\Gamma_{\omega}
\rightarrow \Gamma_{\omega}^{\rm full}(E_{\omega})$.
Integrating over the phase space for the initial and final
hadrons as in Eq.~(\ref{eq:decayrate}) sums over contributions from all
kinematically allowed squared four momentum for the omega.  A broader
total width in the (denominator of the) propagator will naturally
reduce the rate for producing photons.
In physical terms, the propagating omega from which radiation
originates can scatter with the strongly interacting
matter and is therefore no longer as free to decay radiatively.
Modified rates are compared in Fig.~\ref{fig:broadrate} with those
using the vacuum width.  The reduction in photon production is more
pronounced at larger temperatures as expected but for
temperature 100--200 MeV, the modified rate is comparable to
$\pi\rho\rightarrow \pi\gamma$ at its peak.  However,
it is no longer significantly larger than the other axial-vector
decays considered here.

To more fully appreciate the relative importance of radiative decay
Fig.~\ref{fig:last} is shown.  In it, the sum of $a_{1}$, $b_{1}$ and
$K_{1}$ decays is compared at $T=200$ MeV with the dominant hadronic
scattering contributions $\pi\rho\rightarrow \pi\gamma$ and
$\pi\pi\rightarrow \rho\gamma$ as well as the decay
$\rho\rightarrow \pi\pi\gamma$ taken from Ref.~\cite{song93}.  The $b_{1}$
results are those from Fig.~\ref{fig:broadrate} which include a
collisional broadened width for the omega.
Radiative decays contribute more than scattering for photon energies
0.4--0.75 GeV.  Then for more energetic photons they are less important.
If one is merely concerned with
the overall order of magnitude of the photon energy
spectrum, these are clearly not so important.  At some level in more
detailed analyses of the energy spectrum, these axial-vector
radiative decays do become important.

\section{Concluding Remarks}
\label{sec:conclusion}

Mechanisms for photon production in hot hadronic matter are numerous.
Most of them involve species whose abundances are too low or whose scattering
rates or decay rates are too small to compete with pion and rho
meson processes.  There are some with abundances and relevant rates
that do in fact compete for limited range in photon energies.  Namely,
radiative decay of the heavier axial-vector mesons $a_{1}$, $b_{1}$
and $K_{1}$ are relatively important.  Contributions
from $a_{1}\rightarrow \pi\gamma$ and  $K_{1}\rightarrow
K\gamma$ are comparable to $\omega\rightarrow
\pi^{0}\gamma$ with similar photon energy dependence.  Contribution
of $b_{1}\rightarrow \pi\pi^{0}\gamma$ is as large as
$\pi\rho\rightarrow \pi\gamma$ near its peak which occurs at photon
energy 0.5 GeV.
However, for photon energies 0.4 GeV and less,
other processes like $\pi\pi\rightarrow \rho\gamma$
and $\rho\rightarrow \pi\pi\gamma$ become dominant\cite{jkplds,song93}.
Other heavy mesons do not contribute as strongly as the three
axial-vector radiative decays mentioned above.  The $\pi\rho$ and $K\rho$
decay channels of $a_{1}$ and $K_{1}$ are very strong which result in
rather large coupling constants.
By vector-meson dominance, the electromagnetic-decay coupling constants
are also relatively large.  Similar remarks can be made about the
$b_{1}$, but in addition it is also exceptional since one of its most likely
decay products is $\omega$.  The $\omega$'s electromagnetic decay
rate is rather large---a partial width of 0.7 MeV.  Other non-strange
and strange heavy mesons which decay to omega do so with much smaller
rates and furthermore, will not be nearly so abundant at temperatures
100--200 MeV.

Thermalization is assumed for all hadronic species in this study.
For pions and rho mesons this is quite reasonable but for heavier
hadrons it is not so clear.  It may well be that the axial-vectors
considered here do not thermalize.  Their average dynamics could
be different resulting in different photon production.
Modification due to such effects would be an interesting
pursuit.

The method of collisional broadening included in Sec.~\ref{sec:broadening}
is somewhat simplistic.  One should really compute a finite temperature
propagator and self-energy for the omega to two-loop order.  The presence
of matter modifies the width by modifying the imaginary part
of the self-energy.  It also modifies the pole position of the
propagator, i.e. gives the omega an effective mass,  by introducing a
real part to the self-energy.
Such a calculation would have appeal from the point of view of theory
that the real and imaginary parts would be computed consistently.
What has been done in the present work
represents only a simple, first approximation to the effect.

Dilepton production via hadronic decay is limited, of course, to
invariant masses for the pair at most equal to the mass of the parent
hadron less any final state hadrons.  Therefore, higher mass pairs can
only come from very massive hadrons.  One might even consider charmed
non-strange and strange
mesons, but even at these temperatures they would be rarely produced
indeed.  On the other hand, for intermediate dilepton invariant
masses the results presented
here lead naturally to questions about the role of axial vectors,
the $b_{1}$ in particular, on the invariant mass spectrum of lepton
pairs.

\section*{Acknowledgement}

This work was supported by the National
Science Foundation under grant number PHY-9403666.

\begin{figure}
\caption{One-loop contribution to the photon self-energy
including $a_{1}$ and $K_{1}$ axial-vector mesons (a) and
two-loop contribution containing the $\omega$---which further splits
into a $\pi b_{1}$ loop in (b).}
\label{fig:self-energy}
\end{figure}
\begin{figure}
\caption{Diagrams for radiative single-hadron
decays $a_{1}\rightarrow \pi\gamma$ and $K_{1}\rightarrow K\gamma$
(a) and two-pion radiative decay of the $b_{1}$ in (b).}
\label{fig:decays}
\end{figure}
\begin{figure}
\caption{Thermal photon production via axial-vector radiative
decay at $T$ = 200 MeV.}
\label{fig:three}
\end{figure}
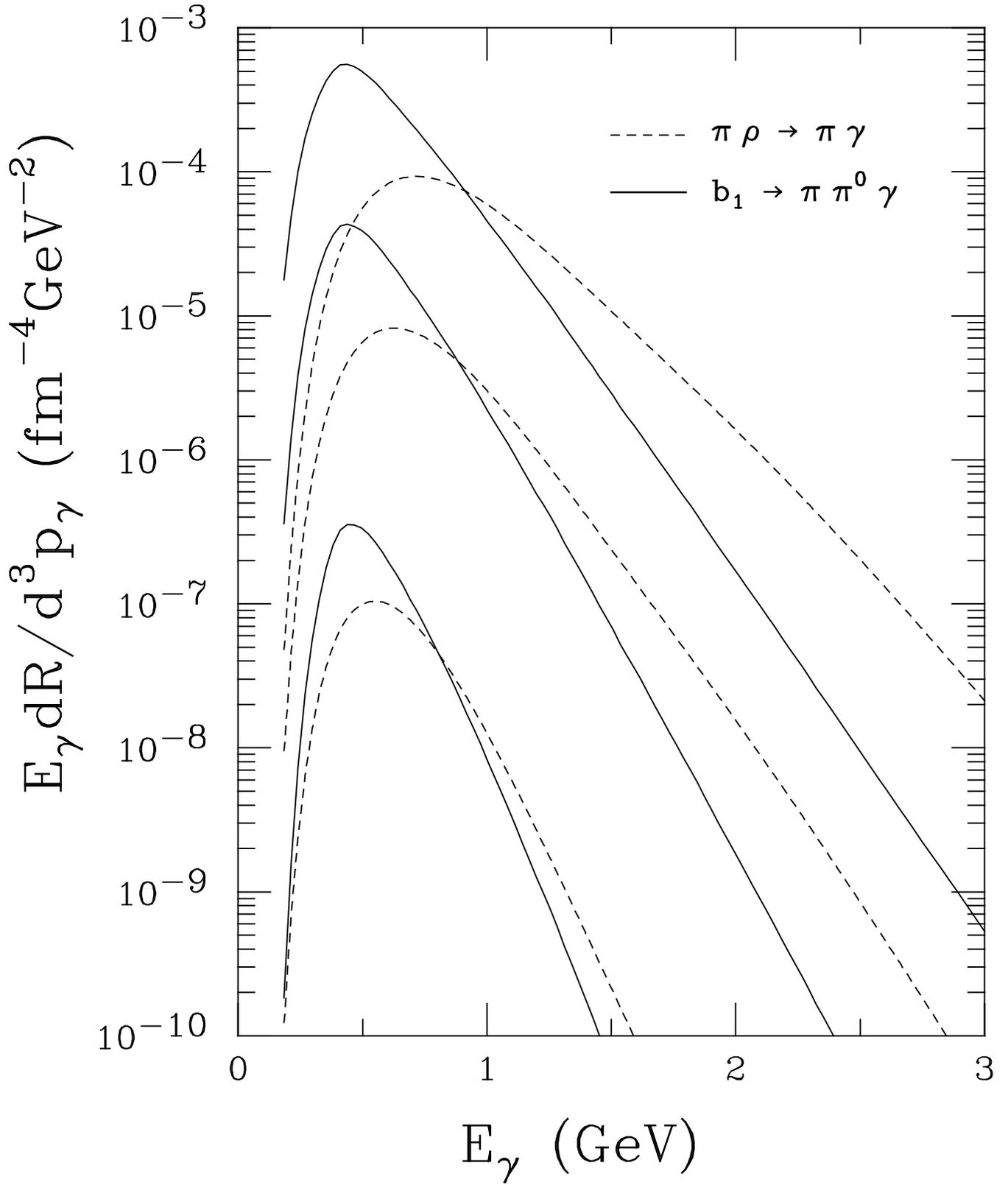
\begin{figure}
\caption{Radiative decay of $b_{1}$ mesons at three temperatures
100, 150 and 200 MeV.  The $b_{1}\rightarrow \pi\pi^{0}\gamma$
results (solid curves) are compared with
$\pi\rho\rightarrow\pi\gamma$ contributions (dashed curves) from
Ref. \ \protect\cite{song93} .}
\label{fig:four}
\end{figure}
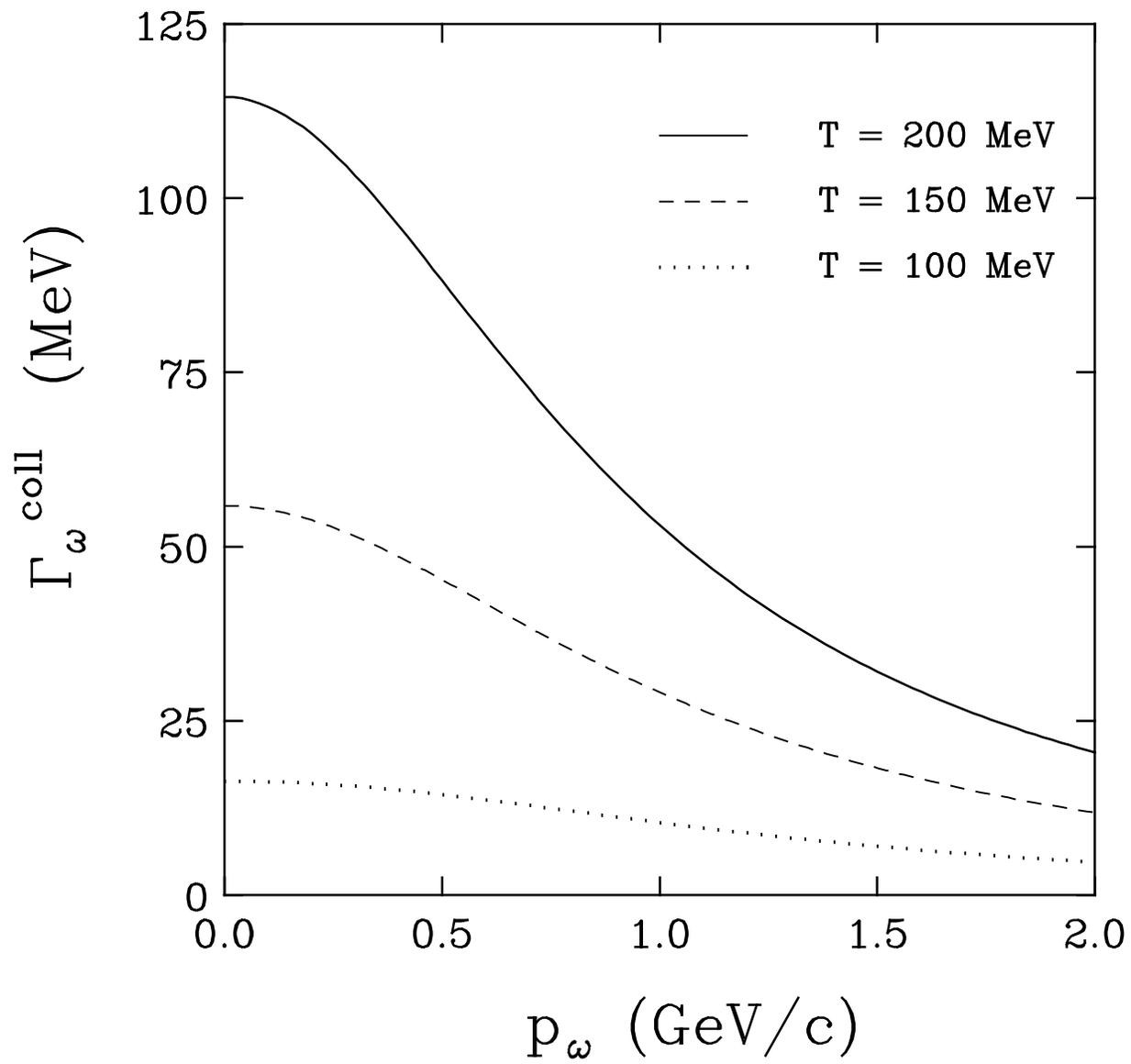
\begin{figure}
\caption{Collisional broadening of the $\omega$ through scattering
with pions as a function of the $\omega$ momentum.  Results for
three temperatures are shown $T$ = 100, 150 and 200 MeV.}
\label{fig:collwidth}
\end{figure}
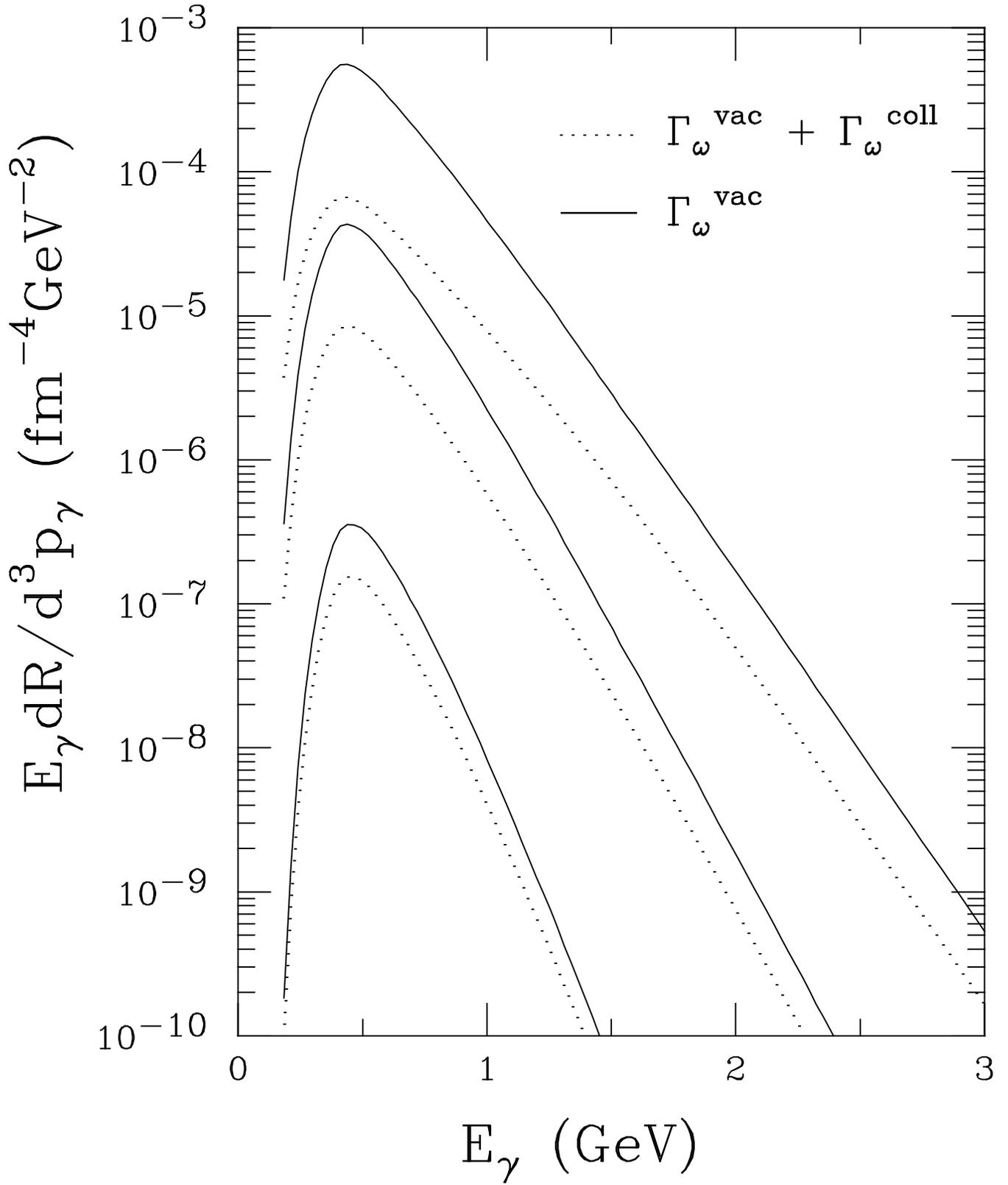
\begin{figure}
\caption{Comparison of $b_{1}\rightarrow \pi \pi^{0}\gamma$ using
the vacuum $\omega$ width (solid curves) and the rate using
a collisional broadened width (dotted curves).  Three temperatures
are again shown, $T$ = 100, 150 and 200 MeV.}
\label{fig:broadrate}
\end{figure}
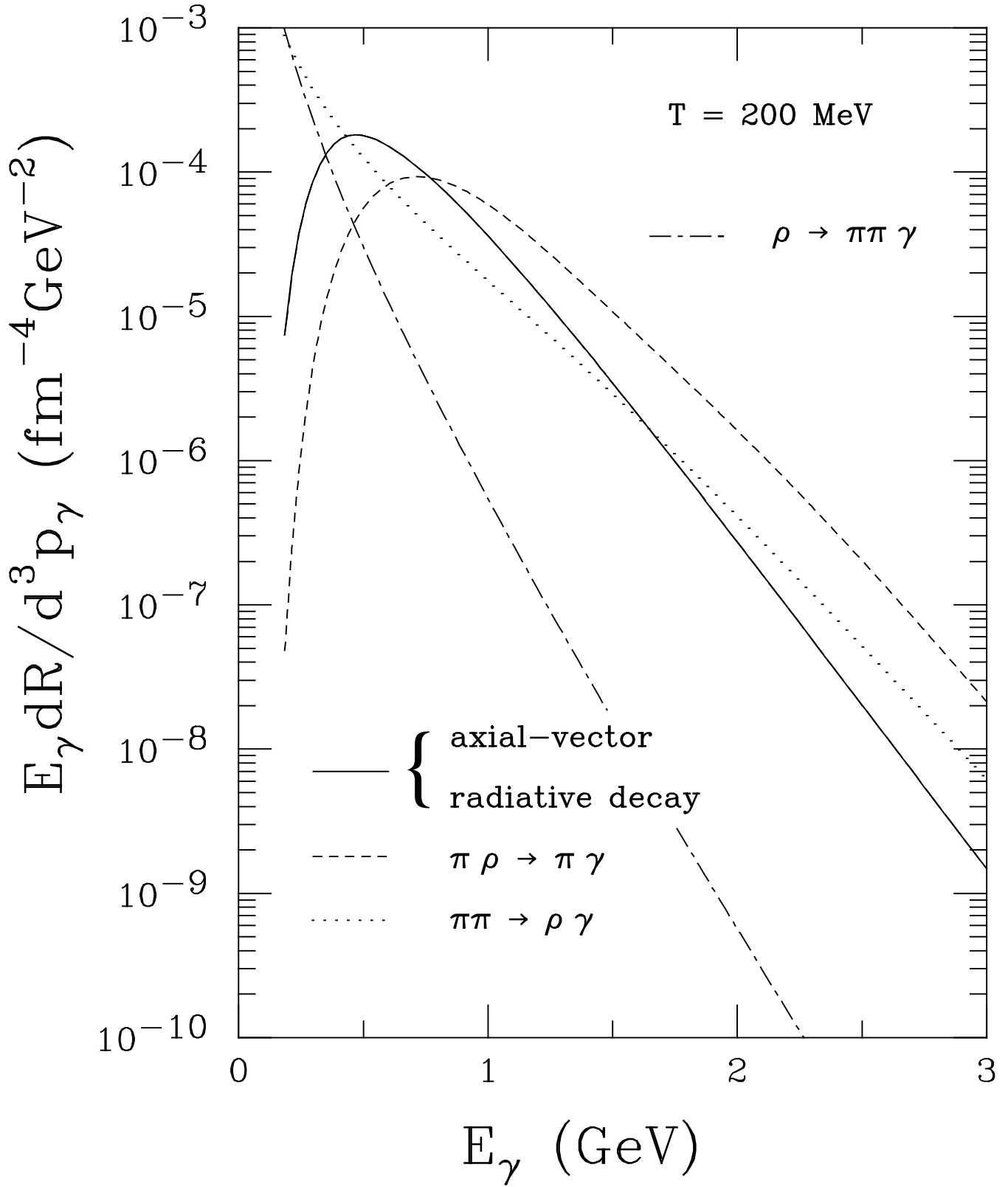
\begin{figure}
\caption{Thermal production rate from the sum of three axial-vector
radiative decays (solid curve) as compared with $\pi\rho\rightarrow \pi\gamma$
(dashed curve), $\pi\pi\rightarrow \rho\gamma$ (dotted curve) and
$\rho\rightarrow \pi\pi\gamma$ (dot-dashed curve) at $T$ = 200 MeV.}
\label{fig:last}
\end{figure}

\begin{references}
\bibitem{khsp2}The $\rho(770)$ and $K^*(892)$ results are presented in
\cite{khsp}, while the same approximations are used to calculate
the numbers for $b_{1}(1235)$ and $K_{1}(1270)$.
\end{references}
\end{document}